\documentclass
[aps,showpacs,prd,onecolumn,amsmath,amssymb]{revtex4-1}
\usepackage{color}
\usepackage{graphics}
\usepackage{epsfig}
\usepackage{amsmath}
\usepackage{amssymb}
\usepackage{amsthm}
\usepackage{slashed}
\usepackage[mathscr]{eucal}

\newcommand{\bea}{\begin{eqnarray}}
\newcommand{\eea}{\end{eqnarray}}
\newcommand{\beas}{\begin{eqnarray*}}
\newcommand{\eeas}{\end{eqnarray*}}

\begin{document}

\preprint{APS/123-QED}

\title{Photon Self-Energy and Electric Susceptibility in a Magnetized Three-flavor Color Superconductor}

\author{Bo Feng}
 \email{bfeng@utep.edu}
\author{Efrain J. Ferrer}
 \email{ejferrer@utep.edu}
\author{Vivian de la Incera}
 \email{vincera@utep.edu}
\affiliation{%
 Department of Physics, University of Texas at El Paso, 500 W. University Ave., El Paso, TX 79968, USA
}%



\date{\today}

\begin{abstract}
We study the photon self-energy for the in-medium photon in a three-flavor color superconductor in the presence of a magnetic field.  At strong magnetic field, the quark dynamics becomes $(1+1)$-dimensional and the self-energy tensor only has longitudinal components. In this approximation there is no Debye or Meissner screenings at zero temperature, but the electric susceptibility is nonzero and highly anisotropic. In the direction transverse to the applied field, the electric susceptibility is the same as in vacuum, while in the longitudinal direction it depends on the magnitude of the magnetic field. Such a behavior is a realization in cold-dense QCD of the magnetoelectric effect, which was first discovered in condensed matter physics. The magnetic permeability remains equal to that in vacuum for both transverse and longitudinal components. We discuss the importance of the Pauli-Villars regularization to get meaningful physical results in the infrared limit of the polarization operator. We also find the covariant form of the polarization operator in the reduced (1+1)-D space of the lowest Landau level and proves its transversality.

\end{abstract}

\maketitle



\section{Introduction}

The phase diagram of Quantum Chromodynamics (QCD) has been under intense studies in recent years \cite{phasediagram}. At extremely high densities and low temperatures color superconductivity (CS) has been established as the favored phase \cite{CS}. The only known natural environment where this state of matter may occur is the superdense core of compact stars. The low-temperature/high-density conditions needed for the formation of a CS ground state might be also produced in low-energy RHIC experiments or in the planned heavy-ion collision experiments at Nuclotron-based Ion Collider (NICA) at JNIR and the Facility for Antiproton and Ion Research (FAIR) at GSI  \cite{HIC}.

Both the natural and laboratory circumstances where a CS state could be formed can be accompanied by the presence of strong magnetic fields. Compact stars are usually strongly magnetized objects. In the case of magnetars,  surface magnetic fields can reach values $\sim 10^{14}-10^{15}$ G \cite{TD}. Because of the high electric conductivity of the stellar medium, during the formation of the neutron star the magnetic flux is conserved. Thus, it is reasonable to expect much stronger magnetic fields in the star core where the matter density is much larger. However, the interior magnetic fields of neutron stars are not  accessible to direct observation, so their magnitudes can only be estimated with the help of heuristic methods. Estimates based on macroscopic and microscopic analysis have led to maximum fields in the range $10^{18}-10^{20}$ G, depending on the nature of the inner medium, that is, whether it is formed by neutrons \cite{DS}, or quarks \cite{Ferrer}. On the other hand,  large magnetic fields of the order of $m_\pi^2\sim 10^{18}-10^{19}$ G can be produced in non-central heavy-ion collisions \cite{HIC}.  If, as claimed in  \cite{CME},  nontrivial topological configurations can separate different charged particles, the existing magnetic field could produce a current that may cause observable consequences in the final states. Moreover,  it has been shown that a sufficiently strong magnetic field can significantly influence the CS state \cite{MCFL,Phases,MCFLoscillation,MagneticmomentMCFL,EM,BCS}.

At this point, it is worth to recall that despite the original electromagnetic $U(1)_{em}$ symmetry is broken by the formation of quark Cooper pairs in the Color-Flavor-Locked (CFL) phase \cite{CFL} of CS, a residual $U(1)$ symmetry still remains. The massless gauge field associated with this symmetry is given by the linear combination of the conventional photon and the $8^{th}$ gluon fields $\tilde A_{\mu}=\cos\theta A_{\mu}-\sin\theta G_{\mu}^8$ \cite{CFL, ABR}. The field $\tilde A_{\mu}$ plays the role of an in-medium or rotated electromagnetic field, as the color condensate is neutral with respect to the corresponding rotated charge. Thus, a magnetic field associated to $\tilde A_{\mu}$ can penetrate the color superconductor and not being subject to the Meissner effect.

In the presence of an external magnetic field, a less symmetric realization of the CFL pairing, known as the Magnetic CFL (MCFL) phase \cite{MCFL}, takes place. The MCFL phase has similarities, but also important differences with the CFL phase \cite{MCFL,MCFLoscillation,Phases,MagneticmomentMCFL}.  For example, the ground state has different symmetry and is characterized by two antisymmetric gaps $\Delta$ and $\Delta_B$, instead of just one, as in the regular CFL case \cite{MCFL}. The change of the ground state symmetry of the three-flavor paired quark matter can be traced back to the rotated charges of the quarks that pair and to their interaction with the rotated magnetic field. One should keep in mind that even though all the diquarks are neutral with respect to the rotated electromagnetic charges, they can be formed either by pairs of neutral or charged quarks with opposite rotated charges. As shown in \cite{MCFL}, the gap $\Delta$ only gets contributions from pairs of neutral quarks, while $\Delta_B$ has contributions from both charged and neutral quarks, thus, it can directly feel the background field through the minimal coupling of the charged quarks with $\tilde B$. At fields large enough that only the lowest Landau level (LLL) is occupied, the field substantially modifies the density of states of the charged quarks and the energy gap $\Delta_B$ becomes significantly enhanced by the penetrating field \cite{MCFL}. At moderate magnetic fields (henceforth when we say magnetic field, we actually mean the rotated magnetic field) the energy gaps exhibit oscillations when ${\tilde e}{\tilde B}/\mu^2$ \cite{MCFLoscillation} is varied, owed to the de Haas-van Alphen effect. In the MCFL phase, as well as in the CFL one, the fermion excitations are gapped, and the gluon fields acquire masses thanks to the Meissner-Anderson-Higgs mechanism, but the symmetry breaking that gives rise to MCFL leaves a smaller number of Nambu-Goldstone fields, all of which are neutral with respect to the rotated
electric charge \cite{Phases}. Hence, the MCFL phase behaves as an insulator, as it has no low-energy charged excitations at zero temperature.

The presence of the magnetic field explicitly breaks the spatial rotational symmetry $O(3)$ to the subgroup $O(2)$ of rotations about the axis parallel to the field. In a recent paper \cite{MagneticmomentMCFL}, we proved that  this symmetry reduction has non-trivial consequences for the ground state structure of the MCFL superconductor.  Performing Fierz transformations in a quark system with both Lorentz and rotational $O(3)$ symmetries explicitly broken, we  uncovered new pairing channels that favor the formation of new condensates. Of particular interest is an attractive channel that  leads to a spin-one condensate of Dirac structure $\Delta_M\sim C\gamma_5\gamma^1\gamma^2$.  A gap of this type does not break any symmetry that has not been already broken, so it is not forbidden in principle. This condensate corresponds to an average magnetic moment of the Cooper pairs in the medium.  From a physical point of view, it is natural to expect the formation of this extra condensate in the magnetized system because the diquarks formed by oppositely charged quarks will have a net magnetic moment that may point parallel or antiparallel to the magnetic field.  Diquarks formed by quarks lying on any non-zero Landau level can have magnetic moments pointing in both directions, because each quark in the pair may have both spins. Hence the contribution of these diquarks to the net magnetic moment should tend to cancel out. On the other hand,  diquarks from quarks in the zero Landau level can only have one orientation of their magnetic moment with respect to the field, because the quarks in the pair have only one possible spin projection.  This implies that the main contribution to the new condensate should come from the quarks at the zero Landau level, an expectation that is consistent with  our numerical results \cite{MagneticmomentMCFL}. The new gap was obtained to be negligibly small at weak magnetic fields, where the zero Landau level occupation is not significant. On the other hand, at strong magnetic fields, the condensate became comparable in magnitude to the original condensates, $\Delta$ and $\Delta_B$, of the MCFL ground state \cite{MCFL}, because the majority of the quarks occupy the zero Landau level in that case. Although this new condensate is zero at zero magnetic field, we cannot ignore it even at very small magnetic fields because a self-consistent solution of the gap equations with $\Delta\neq 0$, and $\Delta_B\neq 0$, but  $\Delta_M=0$ is not allowed. This is easy to understand since there is always some occupation of the zero Landau level, as long as the magnetic field is not zero. Thus, once a magnetic field is present,  $\Delta_M$ has to be considered simultaneously with the spin-zero MCFL gaps. The $\Delta_M$ condensate of the MCFL phase shares a few similarities with the dynamical generation of an anomalous magnetic moment recently found in massless QED \cite{AMM-QED}.

In a recent letter \cite{EM}, we reported the peculiar electromagnetic response of the MCFL medium, which exhibits a magnetoelectric (ME) effect. The ME effect is the capacity of certain materials to have an electric polarization that depends on an applied magnetic field and/or a magnetization that depends on an applied electric field. The ME effect has been known in  condensed matter for many years (see Ref. \cite{ME-Rev} for a review), but to the best of our knowledge it had not been previously found in the context of QCD. Neither the CFL \cite{Manuel} nor  the MCFL \cite{EM} phases have Debye screening, because no infrared electric screening can be produced by  diquark condensates that are neutral.

In the present paper we provide the details of the calculations that lead to the photon self-energy in the MCFL phase with two scalar and one spin-one diquark condensates and use it to obtain the linear electromagnetic response of the system in the strong magnetic field region.  As it will be shown, in this region the MCFL phase exhibits an anisotropic electric polarization, manifested by a different dielectric behavior in the directions parallel and transverse to the magnetic field. Thus, the MCFL phase behaves at strong magnetic fields as a highly anisotropic dielectric medium.

The paper is organized as follows. In Sec. II, the expression for the one-loop polarization operator in the presence of a rotated magnetic field is obtained in momentum space using Ritus's method. In Sec. III, we calculate the full quark propagator for the MCFL phase at strong magnetic field. Using this propagator  and Pauli-Villars regularization,  the polarization operator tensor in the strong-field limit is obtained in Sec. IV. In Sec. V, we obtain the electric permittivity and magnetic permeability of the MCFL phase from the different components of the polarization tensor at strong magnetic field in the infrared limit, $p_0=0, p_3\rightarrow 0$. In Sec. VI, the covariant structure of the polarization operator tensor in momentum space is obtained and the gauge invariance in the strong-field region is verified. Our conclusions are given in Sec. VII. Finally, in Appendix A, the transversality of the in-medium photon polarization operator in the strong-field approximation is independently proved on general grounds.


\section{The photon self-energy at $\widetilde{B}\neq 0$}

Let us consider  a NJL three-flavor model of massless quarks with Lagrangian density
\begin{equation}
{\cal L}=\bar\psi(i\slashed\partial+\mu\gamma_0)\psi-G(\bar\psi\Gamma^a_\mu\psi)(\bar\psi\Gamma_a^\mu\psi).
\label{lagrangian}
\end{equation}
and baryon chemical potential $\mu$. The ground state of this theory is the CFL color superconducting phase.  In the presence of an external magnetic field the ground state of the theory (\ref{lagrangian}) changes into the less symmetric MCFL ground state \cite{MCFL}. As mentioned in the Introduction, the explicit breaking of the rotational symmetry by the external magnetic field allows for new attractive channels and, as shown in  \cite{MagneticmomentMCFL}, the MCFL ground state turns out to have three independent gaps: two scalar ($\Delta$ and $ \Delta_B$) and one spin-one ($\Delta_M$).  All the diquarks in the MCFL phase are neutral with respect to the rotated electromagnetism, so the in-medium electromagnetic  gauge symmetry that exits in the CFL phase also remains intact in the MCFL one.

In this paper we are interested in investigating the static electromagnetic linear response of  the MCFL superconductor in the region of strong magnetic fields. For that, we need to calculate the rotated photon self-energy (i.e. the polarization operator for the in-medium (rotated) photon) in the presence of the external rotated magnetic field, $\widetilde{B}$.  Only the quarks with non zero rotated charge can contribute to the polarization operator. Then, the polarization operator for the rotated photon in coordinate space is
\begin{equation}
\Pi^{\mu\nu}(x,y)=\frac{{\tilde e}^2}{2}\sum_{{\tilde Q}=\pm}Tr\left[\Gamma^\mu{\cal S}_{(\tilde Q)}(x,y)\Gamma^\nu{\cal S}_{(\tilde Q)}(y,x)\right],
\label{selfenergy} 
\end{equation}
with
\begin{equation}
{\cal S}_{(\tilde Q)}(x,y)=\left(
\begin{array}{cc}
G^+_{(\tilde Q)}(x,y) & \Xi^-_{(\tilde Q)}(x,y)\\
\Xi^+_{(\tilde Q)}(x,y) & G^-_{(\tilde Q)}(x,y)
\end{array}
\right),
\label{fullpropagator}
\end{equation}
 the Nambu-Gorkov (NG) full quark propagator of the charged quarks and
\begin{equation}
\Gamma^\mu=\left(
\begin{array}{cc}
{\tilde Q}\gamma^\mu & 0\\
0 & {\tilde Q}\gamma^\mu
\end{array}
\right),\label{vetex}
\end{equation}
the in-medium electromagnetic vertex. In the above formulas $\widetilde{e}$ is the rotated electromagnetic coupling, $\tilde{Q}=\pm $ is the eigenvalue of the rotated charge operator $\mathbb{\tilde Q}=\mathbb{Q}\times 1+1\times \mathbb{T}_8/\sqrt{3}$  with $\mathbb{Q}={\rm diag}(-1/3,-1/3,2/3)$ for $(s,d,u)$ flavors, and $\mathbb{T}_8={\rm diag}(-1/\sqrt{3},-1/\sqrt{3},2/\sqrt{3})$ for $(b,g,r)$ colors  (for details see \cite{MagneticmomentMCFL}).  Notice that the polarization operator (\ref{selfenergy}) gets separate contributions from the positively and negatively charged NG spinors
\begin{equation}
\Psi_+=\left(
\begin{array}{c}
\psi_{(+)}\\
\psi_{(-)C}
\end{array}
\right),
\hspace{0.5cm}
\Psi_-=\left(
\begin{array}{c}
\psi_{(-)}\\
\psi_{(+)C}
\end{array}
\right),
\end{equation}
with subindex ($\pm$) indicating the sign of the rotated electric charge \cite{MCFL}.

The diagonal and off-diagonal elements in (\ref{fullpropagator}) are given respectively by
\begin{equation}
G^{\pm}_{(\tilde Q)}(x,y)=\left\{[G_{(\tilde Q)0}^{\pm}]^{-1}(x,y)-\Phi^{\mp}_{(\tilde Q)}G_{(\tilde Q)0}^{\mp}(x,y)\Phi^\pm_{(\tilde Q)}\right\}^{-1}
\end{equation}
\begin{equation}
\Xi^{\pm}_{(\tilde Q)}(x,y)=-G_{(\tilde Q)0}^{\mp}(x,y)\Phi^\pm _{(\tilde Q)}G^\pm_{(\tilde Q)}(x,y)
\end{equation}
with  $[G^\pm_{(\tilde Q)0}]^{-1}(x,y)=(i\gamma^\mu\partial_\mu+{\tilde e}{\tilde Q}\gamma^\mu \widetilde{A}_\mu\pm\mu\gamma_0)\delta^4(x-y)$ the bare inverse propagators for quarks ($+$) and antiquarks ($-$), and
$\Phi^\pm_{(\tilde Q)}$  the gap matrix whose explicit form will be given later.

We can use the so-called Ritus's method \cite{Ritus, Leung, MCFL} to transform the polarization operator (\ref{selfenergy}) to momentum space. In this approach, the transformation is carried out by the eigenfunctions ${\bf E}^{l(\widetilde{Q})}_p(x)$, which correspond to the asymptotic states of the charged fermions in a uniform magnetic field. Assuming a magnetic field pointing along the positive $x_3$-direction, the ${\bf E}^{l(\widetilde{Q})}_p(x)$ can be written as
\begin{equation}
 {\bf E}^{l(+)}_p(x)=E^{l(+)}_p(x)\Delta(+)+E^{l-1(+)}_p(x)\Delta(-),
\end{equation}
\begin{equation}
{\bf E}^{l(-)}_p(x)=E^{l-1(-)}_p(x)\Delta(+)+E^{l(-)}_p(x)\Delta(-),
\label{epfunction}
\end{equation}
Here $\Delta(\sigma)=(1+i\sigma\gamma^1\gamma^2)/2$ are spin up $(\sigma=+)$ and down $(\sigma=-)$ projectors that satisfy the following relations
\begin{equation}
\Delta(\sigma)+\Delta(-\sigma)=1, \quad \Delta(\sigma)\Delta(-\sigma)=0, \quad \Delta^2(\sigma)=\Delta(\sigma), \quad \Delta(\sigma)\gamma_\|^\mu=\gamma_\|^\mu \Delta(\sigma), \quad \Delta(\sigma)\gamma_{\perp}^\mu=\gamma_{\perp}^\mu \Delta(-\sigma),
\label{deltasproperties}
\end{equation}
where $\|$ refers to the longitudinal components (0,3) and $\perp$ to the transverse ones (1,2) with respect to the direction of the magnetic field. The non-negative integer $l=0,1,2,...$ denotes the Landau levels and the eigenfunctions
\begin{equation}
E^{l(\pm)}_p(x)={\cal N}_le^{-i(p_0x^0+p_2x^2+p_3x^3)}D_l(\rho_{(\pm)}),
\end{equation}
have normalization constant ${\cal N}_l=(4\pi |{\tilde e}{\tilde B}|)^{1/4}/\sqrt{l!}$. The functions $D_l(\rho_{(\pm)})$ are the parabolic cylinder functions of argument $\rho_{(\pm)}=\sqrt{2|{\tilde e}{\tilde B}|}(x_1\pm p_2/{\tilde e}{\tilde B})$ and index given by Landau level numbers.

The ${\bf E}^{l(\pm)}_p$ functions satisfy the equation
\begin{equation}
\int d^4x{\bar{\bf E}}^{l(\pm)}_p(x){\bf E}^{l^\prime(\pm)}_{p^\prime}(x)=(2\pi)^4{\hat\delta}^{(4)}(p-p^\prime)\Pi_{(\pm)}(l),
\end{equation}
with ${\bar{\bf E}}^l_p\equiv\gamma_0({\bf E}^l_p)^\dagger\gamma_0$, ${\hat\delta}^{(4)}(p-p^\prime)=\delta^{ll^\prime}\delta(p_0-p_0^\prime)\delta(p_2-p_2^\prime)\delta(p_3-p_3^\prime)$, and
$
\Pi_{(\pm)}(l)=\Delta(\pm)\delta^{l0}+(1-\delta^{l0}).
$

The ${\bf E}^{l(\pm)}_p$ functions (\ref{epfunction}) satisfy
\begin{equation}
(i\partial_\mu\pm{\tilde e}{\tilde A}_\mu)\gamma^\mu {\bf E}^{l(\pm)}_p(x)={\bf E}^{l(\pm)}_p(x)(\gamma^\mu{\bar p}_\mu^{(\pm)}),
\label{eigenequation}
\end{equation}
with ${\bar p}^{(\pm)}$ given by
\begin{equation}
{\bar p}^{(\pm)}=(p_0,0,\pm\sqrt{2|{\tilde e}{\tilde B}|l},p_3).
\end{equation}

Since photons are neutral, the transformation to momentum space of the LHS of (\ref{selfenergy}) can be carried out by the usual Fourier transform, where we assume translational invariance of the photon self-energy in the presence of a uniform magnetic field
\begin{equation}
(2\pi)^4\delta^{(4)}(p-p^\prime)\Pi^{\mu\nu}(p)=\int d^4xd^4x^\prime e^{-i(p\cdot x-p^\prime\cdot x^\prime)}\Pi^{\mu\nu}(x,x^\prime).
\end{equation}
In terms of the ${\bf E}_p$ functions, the charged quark propagator can be written as
\begin{equation}
S_{(\tilde Q)}(x,x^\prime)=\int\hspace{-0.5cm}\sum\frac{d^4p}{(2\pi)^4}E^{l{(\tilde Q)}}_p(x)\Pi_{(\tilde Q)}(l){\tilde S}_{(\tilde Q)}^l({\bar p}^{(\tilde Q)}){\bar E}^{l{(\tilde Q)}}_p(x^\prime),
\label{S-x}
\end{equation}
where
\begin{equation}\label{full-prog-matrix}
{\tilde S}_{(\tilde Q)}^l({\bar p}^{(\tilde Q)})=\left(
\begin{array}{cc}
{G^+_{(\tilde Q)}}^{l}(\bar p^{(\tilde Q)}) & {\Xi^-_{(\tilde Q)}}^{l}(\bar p^{(\tilde Q)})\\
{\Xi^+_{(\tilde Q)}}^{l}(\bar p^{(\tilde Q)}) & {G^-_{(\tilde Q)}}
^{l}(\bar p^{(\tilde Q)})
\end{array}
\right).
\end{equation}
and  $~\int\hspace{-0.4cm}\sum\frac{d^4p}{(2\pi)^4}\equiv \sum_l\int\frac{dp_0dp_2dp_3}{(2\pi)^4}$. The matrix $\Pi_{(\tilde Q)}(l)$ in (\ref{S-x}) ensures that only one spin projection contributes at the LLL.

Hence, in momentum space the rotated photon self-energy reads
\begin{widetext}
\begin{align}
\nonumber(2\pi)^4\delta^{(4)}(p-p^\prime)\Pi^{\mu\nu}(p)=&\frac{{\tilde e}^2}{2}\sum_{\tilde Q}\int d^4xd^4x^\prime\int\hspace{-0.5cm}\sum\frac{d^4k}{(2\pi)^4}\int\hspace{-0.5cm}\sum\frac{d^4q}{(2\pi)^4}e^{-i(p\cdot x-p^\prime\cdot x^\prime)}\\
&\times{\rm Tr}\left[\gamma^\mu {\bf E}^{l{(\tilde Q)}}_k(x)\Pi_{(\tilde Q)}(l){\tilde S}_{{(\tilde Q)}}^l({\bar k}^{(\tilde Q)}){\bar{\bf E}}^{l{(\tilde Q)}}_k(x^\prime)\gamma^\nu {\bf E}^{{l^\prime}{(\tilde Q)}}_q(x^\prime)\Pi_{(\tilde Q)}({l^\prime}){\tilde S}_{(\tilde Q)}^{l^\prime}({\bar q}^{(\tilde Q)}){\bar{\bf E}}^{{l^\prime}{(\tilde Q)}}_q(x)\right].
\label{selfenergyinmomentum}
\end{align}
\end{widetext}

The integrations in the spatial coordinates give \cite{Leung}
\begin{align}
\nonumber\int d^4x {\bar{\bf E}}^{{l^\prime}({\tilde Q})}_q(x)\gamma^\mu{\bf E}^{l({\tilde Q})}_k(x)e^{-ip\cdot x}=&(2\pi)^4\delta^{(3)}(q+p-k)e^{-ip_1(k_2+q_2)/{2{\tilde e}{\tilde B}}}e^{-{p}^2_\perp/2}\\
&\times\sum_{\sigma, \sigma^\prime}\frac{1}{\sqrt{n!n^\prime !}}e^{isgn({\tilde e}{\tilde B})(n-n^\prime)\varphi}J_{nn^\prime}({\hat p}_\perp)\Delta(\sigma)\gamma^\mu\Delta(\sigma^\prime),
\label{integralone}
\end{align}
and similarly,
\begin{align}
\nonumber\int d^4x^\prime {\bar{\bf E}}^{l({\tilde Q})}_k(x^\prime)\gamma^\nu{\bf E}^{{l^\prime}({\tilde Q})}_q(x)e^{ip^\prime\cdot x}=&(2\pi)^4\delta^{(3)}(q+p^\prime-k)e^{ip^\prime_1(k_2+q_2)/{2{\tilde e}{\tilde B}}}e^{-{p}^{\prime 2}_\perp/2}\\
&\times\sum_{{\bar\sigma},{\bar\sigma}^\prime}\frac{1}{\sqrt{{\bar n}!{\bar n}^\prime!}}e^{isgn({\tilde e}{\tilde B})({\bar n}-{\bar n}^\prime)\varphi}J_{{\bar n}{\bar n}^\prime}({\hat p}^\prime_\perp)\Delta({\bar\sigma})\gamma^\nu\Delta({\bar\sigma}^\prime).
\label{integraltwo}
\end{align}
with $n\equiv n({l},\sigma)$, $n^\prime\equiv n(l^\prime,\sigma^\prime)$, ${\bar n}\equiv n(l,{\bar\sigma})$ and ${\bar n}^\prime\equiv n({l^\prime},{\bar\sigma}^\prime)$, defined according to
\begin{equation}
n(l,\sigma)=l+sgn({\tilde e}{\tilde B})\frac{\sigma}{2}-\frac{1}{2},
\end{equation}
The normalized transverse momentum and the polar angle are defined as ${\hat p}_\perp\equiv \sqrt{{\hat p}^2_1+{\hat p}^2_2}$ and $\varphi\equiv \arctan({\hat p}_2/{\hat p}_1)$, respectively, where ${\hat p}_\mu\equiv p_\mu/\sqrt{2{|\tilde e \tilde B|}}$; the delta function
\begin{equation}
\delta^{(3)}(q+p-k)=\delta(q_0+p_0-k_0)\delta(q_2+p_2-k_2)\delta(q_3+p_3-k_3),
\end{equation} 
and
\begin{align}
J_{nn^\prime}({\hat p}_\perp)=\sum_{m=0}^{min(n,n^\prime)}&\frac{n!n^\prime !}{m!(n-m)!(n^\prime-m)!}\left[isgn({\tilde e}{\tilde B}){\hat p}_\perp\right]^{n+n^\prime-2m}.
\end{align}
The presence of the delta functions in (\ref{integralone}) and (\ref{integraltwo}) enable integrations over $k_0, k_2$ and $k_3$ in (\ref{selfenergyinmomentum}) yielding $\delta^{(3)}(p-p^\prime)=\delta(p_0-p_0^\prime)\delta(p_2-p_2^\prime)\delta(p_3-p_3^\prime)$ and the integral over $q_2$ gives rise to $\delta(p_1-p_1^\prime)$, which combined match the delta function on the LHS of (\ref{selfenergyinmomentum}). Thus, we have
\begin{widetext}
\begin{align}
\nonumber\Pi^{\mu\nu}(p)=&\frac{{\tilde e}^2}{2}|{\tilde e}{\tilde B}|\sum_{\tilde Q}\sum_{l,l^\prime}\sum_{[\sigma]}\int\hspace{-0.5cm}\sum\frac{d^3q}{(2\pi)^3}\frac{e^{isgn({\tilde e}{\tilde B})(n-n^\prime+{\bar n}-{\bar n}^\prime)\varphi}}{\sqrt{n!n^\prime!{\bar n}!{\bar n}^\prime!}}e^{-{\hat p}^2_\perp}J_{nn^\prime}({\hat p}_\perp)J_{{\bar n}{\bar n}^\prime}({\hat p}_\perp)\\
&\times{\rm Tr}\left[\Delta(\sigma)\gamma^\mu\Delta(\sigma^\prime)\Pi_{(\tilde Q)}(l){\tilde S}_{(\tilde Q)}^l({\bar p}^{(\tilde Q)}-{\bar q}^{(\tilde Q)})\Delta({\bar\sigma})\gamma^\nu\Delta({\bar\sigma}^\prime)\Pi_{(\tilde Q)}(l^\prime){\tilde S}_{(\tilde Q)}^{l^\prime}({\bar q}^{(\tilde Q)})\right].
\end{align}
\end{widetext}
where $[\sigma]$ means summing over $\sigma, \sigma^\prime, {\bar\sigma}$ and ${\bar\sigma}^\prime$. Because of the factor $e^{-{\hat p}^2_\perp}$ in the integrand, contributions from large values of ${\hat p}_\perp$ are suppressed. This allows us to keep only the terms with the smallest power of ${\hat p}_\perp$ in $J_{nn^\prime}({\hat p}_\perp)$ \cite{Leung}
\begin{equation}
J_{nn^\prime}({\hat p}_\perp)\rightarrow\frac{\left[{\rm max}(n,n^\prime)\right]!}{|n-n^\prime|!}[i{\hat p}_\perp]^{|n-n^\prime|}\rightarrow n!\delta_{n,n^\prime},
\end{equation}
Then, the self-energy becomes
\begin{widetext}
\begin{align}
\nonumber\Pi^{\mu\nu}(p)=&\frac{{\tilde e}^2}{2}|{\tilde e}{\tilde B}|\sum_{\tilde Q}\sum_{l,l^\prime}\sum_{[\sigma]}\int\hspace{-0.5cm}\sum\frac{d^3q}{(2\pi)^3}e^{-{\hat p}^2_\perp}\delta_{n,n^\prime}\delta_{{\bar n},{\bar n}^\prime}\\
&\times{\rm Tr}\left[\Delta(\sigma)\gamma^\mu\Delta(\sigma^\prime)\Pi_{(\tilde Q)}(l){\tilde S}_{(\tilde Q)}^l({\bar p}^{(\tilde Q)}-{\bar q}^{(\tilde Q)})\Delta({\bar\sigma})\gamma^\nu\Delta({\bar\sigma}^\prime)\Pi_{(\tilde Q)}(l^\prime){\tilde S}_{(\tilde Q)}^{l^\prime}({\bar q}^{(\tilde Q)})\right].
\end{align}
\end{widetext}
Taking into account that
\begin{align}
\nonumber\delta_{n,n^\prime}=\delta_{{l^\prime},l}\delta_{\sigma\sigma^\prime}+\delta_{{l^\prime}+\sigma,l}\delta_{-\sigma,\sigma^\prime},
\end{align}
\begin{align}
\delta_{{\bar n},{\bar n}^\prime}=\delta_{{l^\prime},l}\delta_{{\bar\sigma},{\bar\sigma}^\prime}+\delta_{{l^\prime}+{\bar\sigma}^\prime,l}\delta_{-{\bar\sigma}^\prime,{\bar\sigma}},
\end{align}
we can sum in $[\sigma]$ and ${l^\prime}$. Introducing the finite temperature formulation by replacing $\int dq_0 \rightarrow T\sum_{q_0}$, with Matsubara's frequencies, $q_0=(2r+1)\pi T$, $r=0,\pm1,\pm2,...$, we find
\begin{widetext}
\begin{align}
\nonumber\Pi^{\mu\nu}(p)=&\frac{{\tilde e}^2}{2}|{\tilde e}{\tilde B}|\sum_{\tilde Q}\sum_lT\sum_{q_0}\int\frac{dq_3}{(2\pi)^2}e^{-{\hat p}^2_\perp}\left\{{\rm Tr}\left[\Delta(+)\gamma^\mu\Delta(+)\Pi_{(\tilde Q)}(l)){\tilde S}^l_{(\tilde Q)}({\overline{p-q}})\Delta(+)\gamma^\nu\Delta(+)\Pi_{(\tilde Q)}(l){\tilde S}^l_{({\tilde Q})}({\bar q})\right]\right.\\
\nonumber&+{\rm Tr}\left[\Delta(+)\gamma^\mu\Delta(+)\Pi_{(\tilde Q)}(l){\tilde S}^l_{(\tilde Q)}({\overline{p-q}})\Delta(-)\gamma^\nu\Delta(-)\Pi_{(\tilde Q)}(l){\tilde S}^l_{({\tilde Q})}({\bar q})\right]+{\rm Tr}\left[\Delta(-)\gamma^\mu\Delta(-)\Pi_{(\tilde Q)}(l){\tilde S}^l_{(\tilde Q)}({\overline{p-q}})\right.\\
\nonumber&\times\left.\Delta(+)\gamma^\nu\Delta(+)\Pi_{(\tilde Q)}(l){\tilde S}^l_{({\tilde Q})}({\bar q})\right]+{\rm Tr}\left[\Delta(-)\gamma^\mu\Delta(-)\Pi_{(\tilde Q)}(l){\tilde S}^l_{(\tilde Q)}({\overline{p-q}})\Delta(-)\gamma^\nu\Delta(-)\Pi_{(\tilde Q)}(l){\tilde S}^l_{({\tilde Q})}({\bar q})\right]\\
\nonumber&+{\rm Tr}\left[\Delta(+)\gamma^\mu\Delta(-)\Pi_{(\tilde Q)}(l){\tilde S}^l_{(\tilde Q)}({\overline{p-q}})\Delta(-)\gamma^\nu\Delta(+)\Pi_{(\tilde Q)}(l-1){\tilde S}^{l-1}_{({\tilde Q})}({\bar q})\right]+{\rm Tr}\left[\Delta(-)\gamma^\mu\Delta(+)\Pi_{(\tilde Q)}(l){\tilde S}^l_{(\tilde Q)}({\overline{p-q}})\right.\\
&\times\left.\left.\Delta(+)\gamma^\nu\Delta(-)\Pi_{(\tilde Q)}(l+1){\tilde S}^{l+1}_{({\tilde Q})}({\bar q})\right]\right\}.
\label{Pi-mu-nu}
\end{align}
\end{widetext}
Here, for simplicity we defined $\overline{p-q}={\bar p}^{(\tilde Q)}-{\bar q}^{(\tilde Q)}$.

\section{The full quark propagator at strong magnetic field}
In order to calculate (\ref{Pi-mu-nu}), we first need to find the full quark propagator in momentum space ${\tilde S}^l_{(\tilde Q)}
(\bar k)$. Since we are interested in the strong-field region, it is enough to find the LLL propagator ${\tilde S}^{l=0}_{(\tilde Q)}(\bar k)$, since at strong fields all the charged quarks are constrained to the LLL.

The diagonal
\begin{equation}
{G^\pm_{(\tilde Q)}}^l({\bar k})=\left([{G^\pm_{(\tilde Q)0}}^l]^{-1}({\bar k})-\Phi^\mp_{(\tilde Q)}{G^\mp_{(\pm)0}}^l({\bar k})\Phi^\pm_{(\tilde Q)}\right)^{-1}
\label{G-L}
\end{equation}
and off-diagonal
\begin{equation}
{\Xi^\pm_{(\tilde Q)}}^l({\bar k})=-{G^\mp_{(\tilde Q)}}^l({\bar k})\Phi^\pm_{(\tilde Q)}{G^\pm_{(\tilde Q)0}}^l({\bar k}).
\label{Xi-L}
\end{equation}
elements of (\ref{full-prog-matrix}) can be found from the inverse bare propagator $[{G^\pm_{(\tilde Q)0}}^l]^{-1}({\bar k})$ and the gap matrix $\Phi^\pm_{(\tilde Q)}$ of the charged quarks, with $\Phi_{(\tilde Q)}^-=\gamma_0[\Phi^+_{(\tilde Q)}]^\dagger\gamma_0$.

The color-flavor structure of the MCFL gap matrix is \cite{MagneticmomentMCFL}
\begin{widetext}
\begin{equation}
\Phi^{\alpha\beta}_{ij}={\hat\Delta} \epsilon^{\alpha\beta 3}\epsilon_{ij3}+{\hat\Delta_B}(\epsilon^{\alpha\beta 1}\epsilon_{ij1}+\epsilon^{\alpha\beta 2}\epsilon_{ij2})
+{\hat\Delta_M}[\epsilon^{\alpha\beta 1}(\delta_{i2}\delta_{j3}+\delta_{i3}\delta_{j2})+\epsilon^{\alpha\beta 2}(\delta_{i1}\delta_{j3}+\delta_{i3}\delta_{j1})],
\label{generalansatz}
\end{equation}
\end{widetext}
where $\alpha,\beta$ and $i,j$ denote color and flavor indices respectively. Notice that all the coefficients in (\ref{generalansatz}) are actually matrices in Dirac space, for which we defined ${\hat\Delta}={\Delta{\cal C}\gamma_5}$, ${\hat\Delta_B}={\Delta_B{\cal C}\gamma_5}$ and ${\hat\Delta_M}={\Delta_M{\cal C}\gamma_5\sigma_{12}}$. In the subspace of the positively $\psi^{T}_{(+)} =(u_{b},u_{g}, s_{rC},d_{rC})$ and negatively $\psi^{T}_{(-)} =(s_{r},d_{r}, u_{bC},u_{gC})$ charged NG quarks, the corresponding gap matrices $\Phi_{(\pm)}^+$ are proportional to the identity in color and flavor and have Dirac structure
\begin{equation}
\Phi_{(\pm)}^+=(-\Delta_B\pm\Delta_M){\cal C}\gamma_5\Delta(+)+(-\Delta_B\mp\Delta_M){\cal C}\gamma_5\Delta(-).
\label{gapmatrix}
\end{equation}

In the LLL limit ($l=0$), for positively charged quarks we have
\begin{equation}
[{G^\pm_{(+)0}}^{l=0}]^{-1}(k^\parallel)=(k_0\pm\mu)\gamma_0-k_3\gamma_3=\gamma_0\sum_{e=\pm}[k_0-(ek_3\mp\mu)]\Lambda_{{\bf k}_3}^e.
\label{G-LLL}
\end{equation}
and
\begin{equation}
\Phi^\mp_{(+)}[{G^\mp_{(+)0}}^{l=0}]\Phi^\pm_{(+)}=\gamma_0\sum_{e=\pm}\frac{\Lambda_{{\bf k}_3}^e}{k_0-(ek_3\pm\mu)}[(\Delta_B-\Delta_M)^2\Delta(\pm)+(\Delta_B+\Delta_M)^2\Delta(\mp)].
\label{phi-LLL}
\end{equation}
with $\Lambda_{{\bf k}_3}^e=(1+e\gamma_0\gamma_3{\hat{\bf k}}_3)/2$ being the projectors into states of positive $(e=+)$ or negative $(e=-)$ energy. At this point we concentrate in the propagator for the positively charges quarks, because the calculation is totally analogous for the negatively charged quarks. Then, taking into account (\ref{G-LLL}) and (\ref{phi-LLL}), we can easily obtain $G^\pm$ and $\Xi^\pm$ in the LLL, given respectively by
\begin{equation}
{G^\pm_{(+)}}^{l=0}(k^\parallel)=\sum_{e=\pm}\frac{k_0\mp(\mu-ek_3)}{k_0^2-[\epsilon_{k_3}^e]^2}\Lambda_{{\bf k}_3}^{\pm e}\gamma_0\Delta(+)+\sum_{e=\pm}\frac{k_0\mp(\mu-ek_3)}{k_0^2-[{\bar\epsilon}_{k_3}^e]^2}\Lambda_{{\bf k}_3}^{\pm e}\gamma_0\Delta(-),
\label{G-1}
\end{equation}
and
\begin{equation}
{\Xi^\pm_{(+)}}^{l=0}(k^\parallel)=\pm\sum_{e=\pm}\frac{\Delta_0}{k_0^2-[\epsilon_{k_3}^e]^2}\gamma_5\Lambda_{{\bf k}_3}^{\mp e}\Delta(+)\mp\sum_{e=\pm}\frac{\bar\Delta_0}{k_0^2-[{\bar\epsilon}_{k_3}^e]^2}\gamma_5\Lambda_{{\bf k}_3}^{\mp e}\Delta(-).
\label{Xi-1}
\end{equation}
where we defined $\Delta_0=\Delta_B-\Delta_M$ and ${\bar\Delta_0}=\Delta_B+\Delta_M$. The quasiparticle energies are $\epsilon_{k_3}^e\equiv\sqrt{(\mu-ek_3)^2+\Delta_0^2}$ and ${\bar\epsilon}_{k_3}^e\equiv\sqrt{(\mu-ek_3)^2+{\bar\Delta}_0^2}$.


\section{Regularized polarization operator at strong magnetic field}

Now we consider the in-medium photon polarization operator in the strong magnetic field limit, in which all particles are restricted to the LLL only. In this limit, the propagators for both positively and negatively charged particles only differ in the spin projection of their condensates \cite{MagneticmomentMCFL}. Moreover, as we shall see below, the trace over Dirac space is independent of the spin projection and thus the contribution to the polarization operator of the propagators for positively and negatively charged particles are identical. Therefore, it is enough to calculate the contribution from the positively charged particles and multiply it by two.

In the LLL the charged quark propagators only depend on the Dirac structure ${\bar q}_\|\cdot\gamma^\|$, with ${\bar q}_\parallel=(q_0,q_3)$ and $\gamma^\parallel=(\gamma^0,\gamma^3)$. Taking into account the relation (\ref{deltasproperties}), the polarization operator (\ref{Pi-mu-nu}) in the LLL
reduces to
\begin{align}
\Pi_{\mu\nu}(p^{\parallel})={\tilde e}^2|{\tilde e}{\tilde B}|T\sum_{q_0}\int\frac{dq_3}{(2\pi)^2}{\rm Tr}\left[\Delta(+)\gamma^{\parallel}_\mu{\tilde S}^{l=0}_{(+)}({p^\parallel-q^\parallel})\Delta(+)\gamma_\nu^\parallel{\tilde S}^{l=0}_{(+)}({q}^\parallel)\right].
\label{polarizationtensor}
\end{align}

In (\ref{polarizationtensor}), the approximation ${\hat{ p}}_\perp \simeq 0$ was already used, since we are mainly interested in the infrared behavior of the polarization operator, so the exponential $e^{-{{\hat p}^2}_\perp}$ was eliminated (i.e. $e^{-{{\hat p}^2}_\perp}\simeq 1$) in (\ref{Pi-mu-nu}). From (\ref{polarizationtensor}), one can see that the theory has been effectively reduced to $(1+1)$ dimensions. Obviously, the integral (together with the Matsubara sum) in (\ref{polarizationtensor}) is ultraviolet divergent by simple power counting analysis. In order to obtain meaningful results, we need to regularize (\ref{polarizationtensor}). Because the Lorentz symmetry breaking at finite density, dimensional regularization is not appropriate. There is an extra subtlety with using dimensional regularization in this case that concerns to the $\gamma_5$ dependence of the condensates. As known, $\gamma^5$ is an intrinsically four-dimensional object with no analog in the higher dimensions required in the dimensional regularization approach. Thus, in the following, we will use the Pauli-Villars regularization by introducing a counterterm in the polarization operator, which corresponds to free fermions with zero gap but with a large regulator $\Lambda$ playing the role of a mass. At the end, the regulator $\Lambda$ will be set to infinity, eliminating the introduced degrees of freedom that should not contribute to the physical results.  In the Appendix, we prove the legitimacy of this regularization scheme in securing the gauge invariance of the polarization operator of the rotated photon. Under those prescriptions, the regularized polarization operator reads
\begin{align}
\nonumber\Pi_{\mu\nu}(p^{\parallel})_R=&{\tilde e}^2|{\tilde e}{\tilde B}|T\sum_{q_0}\int\frac{dq_3}{(2\pi)^2}\\
&\times\left\{{\rm Tr}\left[\Delta(+)\gamma^{\parallel}_\mu{\tilde S}^{l=0}_{(+)}({p^\parallel-q^\parallel})\Delta(+)\gamma_\nu^\parallel{\tilde S}^{l=0}_{(+)}({q}^\parallel)\right]\right.-\left.{\rm Tr}\left[\Delta(+)\gamma^{\parallel}_\mu{\tilde S}_{\Lambda}({p^\parallel-q^\parallel})\Delta(+)\gamma_\nu^\parallel{\tilde S}_{\Lambda}({q}^\parallel)\right]\right\},
\label{regularized-self-energy}
\end{align}
Here, ${\tilde S}_{\Lambda}({q}^\parallel)$ is the propagator of the regularization counterpart, whose inverse
reads
\begin{equation}
\left[{\tilde{\cal S}}_{\Lambda}(k^\parallel)\right]^{-1}=\left(
\begin{array}{cc}
[G_0^+]^{-1}_\Lambda(k^\parallel) & 0\\
0 & [G_0^-]^{-1}_\Lambda(k^\parallel)
\end{array}
\right),
\end{equation}
with
\begin{equation}
[G^\pm_0]^{-1}_\Lambda(k^\parallel)=k_0\gamma_0-k_3\gamma^3-\Lambda.
\label{Inverse-G}
\end{equation}
Obviously, it is not difficult to calculate the counterpart contribution to the polarization operator. It is actually similar to the (real) photon self-energy in QED with the electron mass replaced by $\Lambda$, for which the results can be found in the literature \cite{QED}. We therefore will not present the detailed calculation of this regularization counterpart and only give the final results.

The trace in (\ref{regularized-self-energy}) extends to Nambu-Gorkov, color-flavor and Dirac space. First, performing the trace over Nambu-Gorkov space, and using the notation (\ref{full-prog-matrix}), we have
\begin{widetext}
\begin{align}
\nonumber\Pi_{\mu\nu}(p^{\parallel})=&{\tilde e}^2|{\tilde e}{\tilde B}|T\sum_{k_0}\int\frac{dk_3}{(2\pi)^2}{\rm Tr}_{c,f,s}\left[\Delta(+)\gamma^{\parallel}_\mu {G^+_{(+)}}^{l=0}(k^\parallel)\Delta(+)\gamma_\nu^\parallel {G^+_{(+)}}^{l=0}(k^\parallel-{p}^\parallel)\right.\\
\nonumber&+\Delta(+)\gamma^{\parallel}_\mu {G^-_{(+)}}^{l=0}(k^\parallel)\Delta(+)\gamma_\nu^\parallel {G^-_{(+)}}^{l=0}(k^\parallel-{p}^\parallel)+\Delta(+)\gamma^{\parallel}_\mu {\Xi^-_{(+)}}^{l=0}(k^\parallel)\Delta(+)\gamma_\nu^\parallel {\Xi^+_{(+)}}^{l=0}(k^\parallel-{p}^\parallel)\\
&+\left.\Delta(+)\gamma^{\parallel}_\mu {\Xi^+_{(+)}}^{l=0}(k^\parallel)\Delta(+)\gamma_\nu^\parallel {\Xi^-_{(+)}}^{l=0}(k^\parallel-{p}^\parallel)\right]+ \Pi_{\mu\nu}(\Lambda).
\label{selfenergyinNG}
\end{align}
\end{widetext}
Where $\Pi_{\mu\nu}(\Lambda)$ corresponds to the contribution of the second term in the RHS of (\ref{regularized-self-energy}).

Let's introduce the mixed representation for the quark propagator in the LLL (see Ref. \cite{Dirk} for a similar representation at zero field)
\begin{equation}
{G^\pm_{(+)}}^{l=0}(\tau,k_3)\equiv T\sum_{k_0}e^{-k_0\tau}{G^\pm_{(+)}}^{l=0}(k^\parallel),
\label{mixedrepre}
\end{equation}
and its inverse
\begin{equation}
{G^\pm_{(+)}}^{l=0}(k^\parallel)\equiv \int_0^{1/T}d\tau e^{k_0\tau}{G^\pm_{(+)}}^{l=0}(\tau,k_3).
\label{mixedrepre-inv}
\end{equation}
Then, after performing the Matsubara sum in terms of a contour integral in the complex $k_0$ plane, we have
\begin{align}
{G^+_{(+)}}^{l=0}(\tau,k_3)=&\sum_e\Lambda^e_{{\bf k}_3}\gamma_0\left\{n_{{\bf k}_3}^e\left[\theta(-\tau)-N(\epsilon_{{\bf k}_3}^e)\right]e^{\epsilon_{{\bf k}_3}^e\tau}\right.-\left.(1-n_{{\bf k}_3}^e)\left[\theta(\tau)-N(\epsilon_{{\bf k}_3}^e)\right]e^{-\epsilon_{{\bf k}_3}^e\tau}\right\},
\label{coutourintegrate1}
\end{align}
\begin{align}
{G^-_{(+)}}^{l=0}(\tau,k_3)=&-\sum_e\gamma_0\Lambda^e_{{\bf k}_3}\left\{n_{{\bf k}_3}^e\left[\theta(\tau)-N(\epsilon_{{\bf k}_3}^e)\right]e^{-\epsilon_{{\bf k}_3}^e\tau}\right.-\left.(1-n_{{\bf k}_3}^e)\left[\theta(-\tau)-N(\epsilon_{{\bf k}_3}^e)\right]e^{\epsilon_{{\bf k}_3}^e\tau}\right\}.
\label{coutourintegrate2}
\end{align}
with $N(x)=(e^{x/T}+1)^{-1}$ and $n_{{\bf k}_3}^e=(\epsilon_{{\bf k}_3}^e+\mu-ek_3)/2\epsilon_{{\bf k}_3}^e$ being respectively the fermion distribution function and occupation numbers of particles ($e=+1$) or antiparticles ($e=-1$) at zero temperature. Substituting (\ref{mixedrepre}) into (\ref{selfenergyinNG}), and making use of the Kubo-Martin-Schwinger relation for fermions
\begin{equation}
{G^\pm_{(+)}}^{l=0}(1/T-\tau,{\bf k}_3)=-{G^\pm_{(+)}}^{l=0}(-\tau,{\bf k}_3)
\end{equation}
and the identity
\begin{equation}
T\sum_{k_0}e^{k_0\tau}=\sum_{m=-\infty}^\infty(-1)^m\delta(\tau-\frac{m}{T}).
\end{equation}
we obtain
\begin{align}
\nonumber T\sum_{k_0}{\rm Tr}_s\left[\Delta(+)\gamma_\mu^\parallel {G^\pm_{(+)}}^{l=0}(k^\parallel)\Delta(+)\gamma_\nu^\parallel {G^\pm_{(+)}}^{l=0}(q^\parallel)\right]=&-\int_0^{1/T}d\tau e^{p_0\tau}{\rm Tr}_s\left[\Delta(+)\gamma_\mu^\parallel {G^\pm_{(+)}}^{l=0}(\tau,k_3)\right.\\
&\times\left.\Delta(+)\gamma_\nu^\parallel\gamma_0 {G^\mp_{(+)}}^{l=0}(\tau,q_3)\gamma_0\right].
\label{46}
\end{align}
Here, we used that $e^{p_0/T}=1$ for bosonic Matsubara frequencies $p_0=-i2r\pi T$ and defined $q_\mu=k_\mu-p_\mu$. Now inserting (\ref{coutourintegrate1})-(\ref{coutourintegrate2}) into (\ref{46}) and integrating over $\tau$, we have
\begin{align}
\nonumber&T\sum_{k_0}{\rm Tr}_s\left[\Delta(+)\gamma_\mu^\parallel {G^+_{(+)}}^{l=0}(k^\parallel)\Delta(+)\gamma_\nu^\parallel {G^+_{(+)}}^{l=0}(q^\parallel)\right]\\
\nonumber=&-\sum_{e,e^\prime}{\cal T}^{+}_{\mu\nu}\left\{\left(\frac{n_{{\bf k}_3}^e(1-n_{{\bf q}_3}^{e^\prime})}{p_0+\epsilon_{{\bf k}_3}^e+\epsilon_{{\bf q}_3}^{e^\prime}}-\frac{n_{{\bf q}_3}^{e^\prime}(1-n_{{\bf k}_3}^e)}{p_0-\epsilon_{{\bf k}_3}^e-\epsilon_{{\bf q}_3}^{e^\prime}}\right)\right.\times\left[1-N(\epsilon_{{\bf k}_3}^e)-N(\epsilon_{{\bf q}_3}^{e^\prime})\right]\\
&+\left.\left(\frac{(1-n_{{\bf k}_3}^e)(1-n_{{\bf q}_3}^{e^\prime})}{p_0-\epsilon_{{\bf k}_3}^e+\epsilon_{{\bf q}_3}^{e^\prime}}-\frac{n_{{\bf k}_3}^en_{{\bf q}_3}^{e^\prime}}{p_0+\epsilon_{{\bf k}_3}^e-\epsilon_{{\bf q}_3}^{e^\prime}}\right)\times\left[N(\epsilon_{{\bf k}_3}^e)-N(\epsilon_{{\bf q}_3}^{e^\prime})\right]\right\},
\label{T-1}
\end{align}
and
\begin{align}
\nonumber&T\sum_{k_0}{\rm Tr}_s\left[\Delta(+)\gamma_\mu^\parallel {G^-_{(+)}}^{l=0}(k^\parallel)\Delta(+)\gamma_\nu^\parallel {G^-_{(+)}}^{l=0}(q^\parallel)\right]\\
\nonumber=&-\sum_{e,e^\prime}{\cal T}^{-}_{\mu\nu}\left\{\left(\frac{(1-n_{{\bf k}_3}^{e})n_{{\bf q}_3}^{e^\prime}}{p_0+\epsilon_{{\bf k}_3}^e+\epsilon_{{\bf q}_3}^{e^\prime}}-\frac{n_{{\bf k}_3}^e(1-n_{{\bf q}_3}^{e^\prime})}{p_0-\epsilon_{{\bf k}_3}^e-\epsilon_{{\bf q}_3}^{e^\prime}}\right)\right.\times\left[1-N(\epsilon_{{\bf k}_3}^e)-N(\epsilon_{{\bf q}_3}^{e^\prime})\right]\\
&+\left.\left(\frac{n_{{\bf k}_3}^en_{{\bf q}_3}^{e^\prime}}{p_0-\epsilon_{{\bf k}_3}^e+\epsilon_{{\bf q}_3}^{e^\prime}}-\frac{(1-n_{{\bf k}_3}^e)(1-n_{{\bf q}_3}^{e^\prime})}{p_0+\epsilon_{{\bf k}_3}^e-\epsilon_{{\bf q}_3}^{e^\prime}}\right)\times\left[N(\epsilon_{{\bf k}_3}^e)-N(\epsilon_{{\bf q}_3}^{e^\prime})\right]\right\},
\label{T-2}
\end{align}
In (\ref{T-1})-(\ref{T-2}) we introduced the notation
\begin{equation}
{\cal T}^\pm_{\mu\nu}={\rm Tr}_s\left[\gamma_0\gamma_\mu^\parallel\Lambda_{{\bf k}_3}^{\pm e}\Delta(+)\gamma_0\gamma_\nu^\parallel\Lambda_{{\bf q}_3}^{\pm e^\prime}\Delta(+)\right].
\label{diagonaltrace}
\end{equation}
The same procedure can be applied to the off-diagonal components in (\ref{selfenergyinNG}) to end up with
\begin{align}
\nonumber&T\sum_{k_0}{\rm Tr}\left[\Delta(+)\gamma^{\parallel}_\mu {\Xi^\mp_{(+)}}^{l=0}(k^\parallel)\Delta(+)\gamma_\nu^\parallel {\Xi^\pm_{(+)}}^{l=0}(q^\parallel)\right]\\
\nonumber=&-\sum_{e,e^\prime}{\cal U}^\pm_{\mu\nu}\frac{\Delta^2}{4\epsilon_{{\bf k}_3}^e\epsilon_{{\bf q}_3}^{e^\prime}}\left\{\left(\frac{1}{p_0+\epsilon_{{\bf k}_3}^e+\epsilon_{{\bf q}_3}^{e^\prime}}-\frac{1}{p_0-\epsilon_{{\bf k}_3}^e-\epsilon_{{\bf q}_3}^{e^\prime}}\right)\right.\times\left[1-N(\epsilon_{{\bf k}_3}^e)-N(\epsilon_{{\bf q}_3}^{e^\prime})\right]\\
&-\left.\left(\frac{1}{p_0-\epsilon_{{\bf k}_3}^e+\epsilon_{{\bf q}_3}^{e^\prime}}-\frac{1}{p_0+\epsilon_{{\bf k}_3}^e-\epsilon_{{\bf q}_3}^{e^\prime}}\right)\times\left[N(\epsilon_{{\bf k}_3}^e)-N(\epsilon_{{\bf q}_3}^{e^\prime})\right]\right\},
\end{align}
with
\begin{equation}
{\cal U}^\pm_{\mu\nu}={\rm Tr}_s\left[\gamma_\mu^\parallel\gamma_5\Lambda_{{\bf k}_3}^{\pm e}\Delta(+)\gamma_\nu^\parallel\gamma_5\Lambda_{{\bf q}_3}^{\mp e^\prime}\Delta(+)\right].
\label{offdiagonaltrace}
\end{equation}
Putting everything together, the polarization operator (\ref{selfenergyinNG}) for rotated photons reads
\begin{widetext}
\begin{eqnarray}
\nonumber\Pi_{\mu\nu}(p^\parallel)=&-&2{\tilde e}^2|{\tilde e}{\tilde B}|\int\frac{dk_3}{(2\pi)^2}\sum_{e,e^\prime}\\
\nonumber&\times&\left\{
\left[\left(\frac{n_1(1-n_2)}{p_0+\epsilon_1+\epsilon_2}-\frac{(1-n_1)n_2}{p_0-\epsilon_1-\epsilon_2}\right)\left(1-N_1-N_2\right)\right.
+\left.\left(\frac{(1-n_1)(1-n_2)}{p_0-\epsilon_1+\epsilon_2}-\frac{n_1n_2}{p_0+\epsilon_1-\epsilon_2}\right)\left(N_1-N_2\right)\right]{\cal T}^{+}_{\mu\nu}\right.\\
\nonumber&+&\left[\left(\frac{(1-n_1)n_2}{p_0+\epsilon_1+\epsilon_2}-\frac{n_1(1-n_2)}{p_0-\epsilon_1-\epsilon_2}\right)\left(1-N_1-N_2\right)\right.
+\left.\left(\frac{n_1n_2}{p_0-\epsilon_1+\epsilon_2}-\frac{(1-n_1)(1-n_2)}{p_0+\epsilon_1-\epsilon_2}\right)\left(N_1-N_2\right)\right]{\cal T}^-_{\mu\nu}\\
\nonumber&+&\left[\left(\frac{1}{p_0+\epsilon_1-\epsilon_2}-\frac{1}{p_0-\epsilon_1-\epsilon_2}\right)(1-N_1-N_2)-\left(\frac{1}{p_0-\epsilon_1+\epsilon_2}-\frac{1}{p_0+\epsilon_1-\epsilon_2}\right)(N_1-N_2)\right]\\
&\times&\left.\left({\cal U}^+_{\mu\nu}+{\cal U}^-_{\mu\nu}\right)\frac{\Delta^2}{4\epsilon_1\epsilon_2}+\Pi_{\mu\nu}(\Lambda)\right\}.
\label{finalselfenergy}
\end{eqnarray}
\end{widetext}
Here some compact notations, similar to those used in \cite{Dirk}, were introduced: $1$ stands for subscript ${\bf k}_3$ and superscript $e$, and $2$ for subscript ${\bf q}_3$ and superscript $e^\prime$. Notice that the prefactor $2$ comes from the trace in color-flavor, since the RHS of (\ref{selfenergyinNG}) is diagonal in color-flavor space.

The remaining trace over Dirac space in (\ref{diagonaltrace}) and (\ref{offdiagonaltrace}) can be easily performed to obtain
\begin{subequations}
\begin{equation}
{\cal T}^\pm_{00}=-{\cal U}^\pm_{00}=\frac{1}{2}(1+ee^\prime {\hat{\bf k}}_3{\hat{\bf q}}_3),
\end{equation}
\begin{equation}
{\cal T}^\pm_{03}={\cal T}^\pm_{30}=-{\cal U}^\pm_{03}={\cal U}^\pm_{30}=\frac{1}{2}(\mp e{\hat{\bf k}}_3\mp e^\prime {\hat{\bf q}}_3),
\end{equation}
\begin{equation}
{\cal T}^\pm_{33}={\cal U}^\pm_{33}=\frac{1}{2}(1+ee^\prime {\hat{\bf k}}_3{\hat{\bf q}}_3).
\end{equation}
\label{trace}
\end{subequations}
Notice that we would end up with the same results if we replace the spin projectors $\Delta(+)$ in (\ref{diagonaltrace}) and (\ref{offdiagonaltrace}) by $\Delta(-)$. That shows that the contribution of the positive and negative charged particles with corresponding spin up ($\Delta(+)$) and spin down ($\Delta(-)$) projectors respectively, to the polarization operator are identical, as we discussed previously. The different components of the rotated-photon polarization operator (\ref{finalselfenergy}) can be explicitly given as
\begin{widetext}
\begin{align}
\nonumber{\Pi_{00}}_R(p^\parallel)=&-{\tilde e}^2|{\tilde e}{\tilde B}|\int_{-\infty}^\infty \frac{dk_3}{(2\pi)^2}\left\{\sum_{e,e^\prime}(1+ee^\prime {\hat{\bf k}}_3{\hat{\bf q}}_3)\right.\times\left[(\frac{1}{p_0+\epsilon_1+\epsilon_2}-\frac{1}{p_0-\epsilon_1-\epsilon_2})(1-N_1-N_2)\frac{\epsilon_1\epsilon_2-\xi_1\xi_2-\Delta_0^2}{2\epsilon_1\epsilon_2}\right.\\
\nonumber&+\left.(\frac{1}{p_0-\epsilon_1+\epsilon_2}-\frac{1}{p_0+\epsilon_1-\epsilon_2})(N_1-N_2)\frac{\epsilon_1\epsilon_2+\xi_1\xi_2+\Delta_0^2}{2\epsilon_1\epsilon_2}\right]+\left[\left(\frac{1}{p_0+E_1+E_2}-\frac{1}{p_0-E_1-E_2}\right)\right.\\
&\times\left.\left.\left(1-N_{E_1}-N_{E_2}\right)\frac{E_1E_2-k_3q_3-\Lambda^2}{E_1E_2}+\left(\frac{1}{p_0-E_1+E_2}-\frac{1}{p_0+E_1-E_2}\right)(N_{E_1}-N_{E_2})\frac{E_1E_2+k_3q_3+\Lambda^2}{E_1E_2}\right]\right\},
\label{00component}
\end{align}
\begin{align}
\nonumber{\Pi_{03}}_R(p^\parallel)=&-{\tilde e}^2|{\tilde e}{\tilde B}|\int_{-\infty}^\infty \frac{dk_3}{(2\pi)^2}\left\{\sum_{e,e^\prime}(e{\hat{\bf k}}_3+e^\prime{\hat{\bf q}}_3)\times\left[\left(\frac{1}{p_0+\epsilon_1+\epsilon_2}+\frac{1}{p_0-\epsilon_1-\epsilon_2}\right)(1-N_1-N_2)\left(\frac{\xi_2}{2\epsilon_2}-\frac{\xi_1}{2\epsilon_1}\right)\right.\right.\\
\nonumber&+\left.\left(\frac{1}{p_0-\epsilon_1+\epsilon_2}+\frac{1}{p_0+\epsilon_1-\epsilon_2}\right)(N_1-N_2)\left(\frac{\xi_1}{2\epsilon_1}+\frac{\xi_2}{2\epsilon_2}\right)\right]+i\left[\left(\frac{1}{p_0+E_1+E_2}+\frac{1}{p_0-E_1-E_2}\right)\right.\\
&\left.\left.\times\left(1-N_{E_1}-N_{E_2}\right)\left(\frac{q_3}{E_2}-\frac{k_3}{E_1}\right)+\left(\frac{1}{p_0-E_1+E_2}-\frac{1}{p_0+E_1-E_2}\right)\left(\frac{q_3}{E_2}+\frac{k_3}{E_1}\right)(N_{E_1}-N_{E_2})\right]\right\},
\end{align}
\end{widetext}
and
\begin{widetext}
\begin{align}
\nonumber{\Pi_{33}}_R(p^\parallel)=&-{\tilde e}^2|{\tilde e}{\tilde B}|\int_{-\infty}^\infty \frac{dk_3}{(2\pi)^2}\left\{\sum_{e,e^\prime}(1+ee^\prime {\hat{\bf k}}_3{\hat{\bf q}}_3)\times\left[\left(\frac{1}{p_0+\epsilon_1+\epsilon_2}-\frac{1}{p_0-\epsilon_1-\epsilon_2}\right)(1-N_1-N_2)\frac{\epsilon_1\epsilon_2-\xi_1\xi_2+\Delta_0^2}{2\epsilon_1\epsilon_2}\right.\right.\\
\nonumber&+\left.\left(\frac{1}{p_0-\epsilon_1+\epsilon_2}-\frac{1}{p_0+\epsilon_1-\epsilon_2}\right)(N_1-N_2)\frac{\epsilon_1\epsilon_2+\xi_1\xi_2-\Delta_0^2}{2\epsilon_1\epsilon_2}\right]-\left[\left(\frac{1}{p_0+E_1+E_2}-\frac{1}{p_0-E_1-E_2}\right)\right.\\
&\times\left.\left.(1-N_{E_1}-N_{E_2})\frac{E_1E_2-k_3q_3+\Lambda^2}{E_1E_2}+\left(\frac{1}{p_0-E_1+E_2}-\frac{1}{p_0+E_1-E_2}\right)(N_{E_1}-N_{E_2})\frac{E_1E_2+k_3q_3-\Lambda^2}{E_1E_2}\right]\right\}.
\label{33component}
\end{align}
\end{widetext}
Here, we used the notation $E_1=\sqrt{k_3^2+\Lambda^2}$, $E_2=\sqrt{q_3^2+\Lambda^2}$ and $\xi_1=ek_3-\mu$, $\xi_2=e^\prime q_3-\mu$.

\section{electric permittivity and magnetic permeability in the strong-field approximation}

In the following, we will consider the zero temperature limit, $T=0$, of the polarization operator components (\ref{00component})-(\ref{33component}), and consequently all the Fermion distribution functions vanish. Moreover, we can drop the contribution of the antiparticles because it is always much smaller than that of the particles. Because we are interested in the infrared behavior of the photon self-energy, we expand (\ref{00component})-(\ref{33component}) in powers of the photon momentum components $p_0$ and $p_3$, up to quadratic terms
\begin{equation}
{\Pi_{00}}_R=-\lim_{\Lambda\rightarrow\infty}\frac{{\tilde e}^2|{\tilde e}{\tilde B}|p_3^2}{6\pi^2}(\frac{1}{\Delta_0^2}+\frac{1}{\Lambda^2})=-\frac{{\tilde e}^2|{\tilde e}{\tilde B}|p_3^2}{6\pi^2\Delta_0^2},
\label{Pi-00}
\end{equation}
\begin{equation}
{\Pi_{33}}_R=-\lim_{\Lambda\rightarrow\infty}\frac{{\tilde e}^2|{\tilde e}{\tilde B}|p_0^2}{6\pi^2}(\frac{1}{\Delta_0^2}-\frac{1}{\Lambda^2})=-\frac{{\tilde e}^2|{\tilde e}{\tilde B}|p_0^2}{6\pi^2\Delta_0^2}.
\label{Pi-33}
\end{equation}
and ${\Pi_{30}}_R={\Pi_{03}}_R\simeq 0$. As we expected, the regulator introduced through the Pauli-Villars regularization scheme does not appear in the final results once we take $\Lambda \rightarrow \infty$.

Because $\Pi_{00}$ has no constant contribution in the infrared limit $p_0=0, p_3\rightarrow 0$, one immediately concludes that there is no Debye screening in the strong-field region, as it was the case at zero field in the CFL phase \cite{Manuel}. This is simply because all quarks are bound within the rotated-charge neutral condensates. There is also no Meissner screening (i.e. $\Pi_{33}$ is zero in the zero-momentum limit), as it should be expected from the remnant $\tilde U(1)$ gauge symmetry. However, the condensates have electric dipole moments and could align themselves in an electric field. Hence, this should modify the dielectric constant of the medium. Since the quadratic term in the effective $\tilde U(1)$ Lagrangian is given by ${\tilde A}_\mu(-p)[D^{-1}_{\mu\nu}(p)+\Pi_{\mu\nu}(p)]{\tilde A}_\nu(p)$, with $D^{-1}$ being the bare rotated photon propagator, the effective action of the $\tilde U(1)$ field in the strong-field region is given by
\begin{equation}
S_{eff}=\int d^4x[\frac{\epsilon_\parallel}{2}{\tilde {\bf E}}_\parallel\cdot {\tilde {\bf E}}_\parallel+\frac{\epsilon_\perp}{2}{\tilde {\bf E}}_\perp\cdot {\tilde {\bf E}}_\perp-\frac{1}{2\lambda_\parallel}{\tilde {\bf H}}_\parallel\cdot {\tilde {\bf H}}_\parallel
-\frac{1}{2\lambda_\perp}{\tilde {\bf H}}_\perp\cdot {\tilde {\bf H}}_\perp],
\label{Action}
\end{equation}
where the separation between transverse and longitudinal parts is due to the $O(3) \rightarrow O(2)$ symmetry breaking produced by the strong magnetic field $\widetilde{B}$. In (\ref{Action}), $\widetilde{E}$, $\widetilde{H}$ are weak electric and magnetic field probes, respectively. In (\ref{Action}) the coefficients $\epsilon$ and $\lambda$ denote the electric permittivity and magnetic permeability of the medium respectively.

From (\ref{Pi-00})-(\ref{Pi-33}) it is straightforward that in the infrared limit the transverse and longitudinal components of the electric permittivity and magnetic permeability become
\begin{equation}
 \lambda_\perp=\lambda_\parallel\simeq 1, \quad
\epsilon_\perp=1, \quad \epsilon_\parallel=1+\chi_{MCFL}^\parallel=1+\frac{{\tilde e}^2|{\tilde e}{\tilde B}|}{6\pi^2\Delta_0^2},
\label{susceptibility}
\end{equation}
where $\chi_{MCFL}^\parallel$ is the longitudinal electric susceptibility. Notice that the longitudinal electric susceptibility is much larger than one because in the strong-magnetic-field limit ${\tilde e}{\tilde B}\gg \Delta_0^2$ \cite{MagneticmomentMCFL}.

Although a static $\tilde U(1)$ charge cannot be completely Debye screened by the $\tilde U(1)$ neutral Cooper pairs, it can still be partially screened along the magnetic field direction because the medium is highly polarizable on that direction. This is due to the existence of Cooper pairs with opposite rotated charges $\widetilde{Q}$ that behave as electric dipoles with respect to the rotated electromagnetism of the MCFL phase. Moreover, the electric susceptibility depends on the magnetic field. When the magnetic field increases in the strong-field region, the susceptibility  becomes smaller, because the coherence length $\xi \sim 1/\Delta_0 $ decreases (i.e. $\Delta_0$ increases) with the field at a quicker rate than $\sqrt{{\tilde e}{\tilde B}}$ \cite{MagneticmomentMCFL}, and the pair's coherence length $\xi$ plays the role of the dipole length.  Hence, with increasing magnetic field the polarization effects weaken in the strong-field region. The tuning of the electric polarization by a magnetic field is what is called in condensed matter physics the magnetoelectric effect. We have discussed this phenomenon in highly magnetized CS in detail in \cite{EM}. From (\ref{susceptibility}), we also see that at strong magnetic fields the medium turns out to be very anisotropic. The fact that the electric permittivity is only modified in the longitudinal direction is due to the confinement of the quarks to the LLL at high enough fields.

\section{Covariant structure and gauge invariance of the polarization tensor }

The photon polarization operator should be gauge invariant. That is, in the strong-field approximation, it should satisfy the transversality condition in the reduced $(1+1)$-D space (${p_\mu}^\parallel\Pi_{\mu\nu}^\parallel(p^\parallel)=0$). As known, the polarization operator tensor can be expanded in a superposition of independent transverse Lorentz tensors. The number of these basic transverse tensors depends on the symmetries of the system under consideration. For example, in vacuum, where the only available tensorial structures are the four-momentum and the metric tensor, there is only one   gauge invariant structure. When a medium is under consideration (i.e. at finite temperature or finite density), since the Lorentz symmetry is broken, there is an additional gauge invariant structure that can be formed by taking into account a new four-vector, the four-velocity of the medium center of mass, $u_\mu$, \cite{Fradkin}. When a magnetic field is applied on that medium, then the structure of the polarization operator is enriched by an additional tensor, $F_{\mu\nu}$. Then, at finite density and in the presence of a magnetic field, there are nine independent gauge-invariant tensorial structures \cite{Shabad}. At strong magnetic field, when the particles are confined to the LLL, due to the fact that the transverse momentum is zero, there is a dimensional reduction leaving only the tensors $g_{\mu\nu}^\parallel, p_\mu^\parallel$ and $u^\|_\mu=(1,0)$ at our disposal. The original nine structures of Ref. \cite{Shabad} now reduce to only two
\begin{equation}
T_{\mu\nu}^{(1)}=(p^\parallel)^2g_{\mu\nu}^\parallel-p_\mu^\parallel p_\nu^\parallel,
\label{structure-1}
\end{equation}
and
\begin{equation}
T_{\mu\nu}^{(2)}=\left[u_\mu^\parallel-\frac{p_\mu^\parallel(u^\parallel\cdot p^\parallel)}{(p^\parallel)^2}\right]\left[u_\nu^\parallel-\frac{p_\nu^\parallel(u^\parallel\cdot p^\parallel)}{p^2}\right].
\label{structure-2}
\end{equation}
Moreover, one can readily check that the two tensors (\ref{structure-1}) and (\ref{structure-2}) are equivalent, which indicates that the rotated-photon polarization operator tensor, at strong magnetic field, only has one independent structure
\begin{equation}
\Pi_{\mu\nu}^\parallel(p^\parallel)=\Pi(p^\parallel,\mu,B)\left[(p^\parallel)^2g_{\mu\nu}^\parallel-p_\mu^\parallel p_\nu^\parallel\right],
\label{Cov-structure}
\end{equation}
with $\Pi(p^\parallel,\mu,B)$ being a scalar coefficient depending on the photon longitudinal momentum, baryonic chemical potential and magnetic field.

Contracting with the tensor $g_{\mu\nu}^\parallel$ both sides of Eq. (\ref{Cov-structure}), we will obtain the scalar function $\Pi(p^\parallel,\mu,B)$ by substituting Eqs. (\ref{G-1})-(\ref{Xi-1}) into Eq. (\ref{polarizationtensor}) and performing the traces in all spaces,
\begin{align}
\nonumber\Pi(p^\|,\mu,B)=&(g^{\mu\nu})^\parallel\Pi_{\mu\nu}^\parallel\\
=&4{\tilde e}^2|{\tilde e}{\tilde B}|T\sum_{q_0}\int\frac{dq_3}{(2\pi)^2}\sum_{e,e^\prime}(1+ee^\prime {\hat {\bf q}_3}{\hat {\bf k}_3})\times\frac{\Delta_0^2}{\left(q_0^2-[\epsilon_{q_3}^e]^2\right)\left(k_0^2-[\epsilon_{k_3}^{e^\prime}]^2\right)}.
\label{trace}
\end{align}
Then, after performing the Matsubara sum, we end up with
\begin{align}
\nonumber\Pi(p^\|,\mu,B)=&-4{\tilde e}^2|{\tilde e}{\tilde B}|\int\frac{dq_3}{(2\pi)^2}\sum_{e,e^\prime}(1+ee^\prime {\hat{\bf q}}_3{\hat{\bf k}}_3)\frac{\Delta_0^2}{4\epsilon_{q_3}^e\epsilon_{k_3}^{e^\prime}}\\
\nonumber&\times\left[\left(\frac{1}{p_0-\epsilon_{q_3}^e-\epsilon_{k_3}^{e^\prime}}-\frac{1}{p_0+\epsilon_{q_3}^e+\epsilon_{k_3}^{e^\prime}}\right)\right.\times(1-N(\epsilon_{q_3}^e)-N(\epsilon_{k_3}^{e^\prime}))\\
&-\left.\left(\frac{1}{p_0+\epsilon_{q_3}^e-\epsilon_{k_3}^{e^\prime}}-\frac{1}{p_0-\epsilon_{q_3}^e-\epsilon_{k_3}^{e^\prime}}\right)\times(N(\epsilon_{q_3}^e)-N(\epsilon_{k_3}^{e^\prime}))\right].
\label{coefficient}
\end{align}
Now, if we expand the components of the polarization operator (\ref{Cov-structure}), after substituting the coefficient $\Pi(p,\mu,B)$ with (\ref{coefficient}), to quadratic orders in powers of $p_0$ and $p_3$, we will obtain exactly the same results as those of Eqs. (\ref{Pi-00}) and (\ref{Pi-33}). Notice that we do not have to regularize the scalar function (\ref{coefficient}), because the divergent terms cancel out when taking the trace in (\ref{trace}). Hence, the medium electric permittivity and magnetic permeability obtained from (\ref{Pi-00})-(\ref{Pi-33}), have a gauge-invariant nature, as they are obtained from the gauge covariant form of the polarization operator tensor (\ref{Cov-structure}).

\section{Concluding remarks}

In this paper we investigate the polarization effects that affect the in-medium photons in the strongly magnetized MCFL phase of CS. With this purpose, we studied the self-energy of the rotated photons associated with the unbroken $\tilde U(1)$ symmetry of the MCFL phase in the presence of a very strong magnetic field. In calculating the one-loop photon polarization operator we used the Pauli-Villars regularization scheme to regularize the diagram's ultraviolet divergencies. The use of this regularization scheme was crucial to get rid of some unphysical results. We call attention that, without such a convenient regularization we would have ended up with a constant Meissner mass, which is of course in contradiction with the remaining $\tilde U(1)$ gauge invariance of the theory. This simply indicates that it is not allowed to take the infrared limit of the Feynman's diagram while it is not regularized. The regularization procedure in this case is however more subtle than in vacuum; where it is possible to use dimensional regularization, hence preserving the gauge symmetry of the polarization operator. In CS, nevertheless, dimensional regularization is not a suitable regularization procedure, since in this high-dense medium, Lorentz-symmetry is broken, and besides, there is an extra complication due to the presence of $\gamma_5$ in the Cooper-pair condensates. As known, $\gamma_5$ is an intrinsically four dimensional object that cannot be generalized to higher dimensions. We thus use the Pauli-Villars regularization scheme. This regularization procedure is not in general gauge invariant for diagrams with internal lines of gauge bosons, since the addition of an auxiliary boson propagator with a mass term (i.e. given by $\Lambda$ in (\ref{Inverse-G})) will break the gauge symmetry. However, for CS in the hard-dense-loop approximation, the diagrams are formed only by internal fermion lines, and then the Pauli-Villars regularization becomes an appropriate gauge invariant approach.

Notwithstanding the difference between the photon polarization in magnetized QED in vacuum and the one we are reporting for the color superconducting medium, we found that there is still one similarity regarding the covariant tensorial structure of the polarization operator in both cases. In the QED case \cite{QED}, the polarization operator in the strong-field limit exhibits only one transverse structure in the subspace of the longitudinal momenta. In the MCFL phase of CS, although for a magnetized dense medium there exist in general nine possible independent covariant structures \cite{Shabad}, in the strong-field limit they reduce to only one in $(1+1)$-D, which is similar to that of QED (see Eq. (\ref{Cov-structure})).  The fact that we can find a covariant transverse representation for the in-medium photon polarization operator is another manifestation of the gauge invariance of the results we are reporting.

We found that there are no Debye and Meissner screenings in the strongly magnetized color superconducting medium. The reason for the lack of Debye screening is simply because all the particles in the medium are condensed in Cooper pairs that are neutral with respect to the rotated charge, and thus, cannot screen a static rotated charge in the medium. On the other hand, since the rotated $\tilde U(1)$ gauge symmetry remains unbroken, we should not expect a different from zero Meissner mass. Nevertheless, the medium has a large electric susceptibility, which indicates that the medium is highly polarizable. A similar conclusion was found for the gluodynamics of the remnant SU(2) symmetry in the 2SC phase of CS at zero magnetic field in Ref. \cite{Son}, and for the rotated electric polarization in the CFL phase at zero magnetic field \cite{Manuel}. Nevertheless, the electric susceptibility we found in the MCFL phase is different from those phases in two aspects. On one hand, it can change with the applied magnetic field, and on the other hand, it is anisotropic. The electric susceptibility in the strongly magnetized MCFL is quantitatively smaller than that of the CFL phase \cite{Manuel} for the same value of the baryonic chemical potential. It is due to the fact that the electric susceptibility is proportional to $|eB|\xi^2$, and the coherent length, $\xi$, decreases as $1/\Delta_0$, and $\Delta_0$ increases with the field at a quicker rate than $\sqrt{eB}$. Hence, as the coherent length of the Cooper pair plays the role of the electric dipole moment length, when the magnetic field increases in the strong field region, the susceptibility becomes smaller. The fact that the electric susceptibility depends on the magnetic field is an effect analogous to the magnetoelectric effect known in condensed matter physics. In \cite{EM} we discussed in details the realization of this magnetoelectric effect in CS, as well as its possible implications for astrophysics.

Comparing our result for the electric susceptibility (\ref{susceptibility}) with that obtained in strongly magnetized QED, $\chi^\|_{QED}=(\alpha/3\pi)(|eB|/ m^2)$ \cite{QED}, we find that the two results are quite similar after the following replacement of the dipole length $1/\Delta_0 \rightarrow 1/m$. Also, there is an extra factor 2 in $\chi_{MCFL}^\parallel$, which comes from the contribution of two electric dipoles associated to the Cooper pairs formed by the charged quarks present in the MCFL phase: $\langle s_r,u_b\rangle$ and $\langle d_r,u_g\rangle$. Nevertheless, there is a substantial difference between the QED electric susceptibility and that of the MCFL phase of CS regarding their dependence on a magnetic field. While in the QED case, $\chi^\|_{QED}$ increases with $B$, since the coherence lenth, $\xi_{QED}\sim 1/m$, does not change with $B$, in the MCFL phase, $\chi_{MCFL}^\parallel$ decreases with $B$, as it was discussed before.
It is also interesting to see that the chromo-electric susceptibility in strongly magnetized QCD, $\chi^\|_{QCD}=(\alpha_s/6\pi)\sum_{q=1}^{N_f}(|eB|/m_q^2)$ \cite{Igor}, also has the same qualitative behavior as $\chi^\|_{QED}$, with a coefficient 6 instead of the 3 found in QED, which is related with the trace in color associated to the gluon vortices.

The anisotropic nature of the electric susceptibility in the strongly magnetized MCFL phase, together with the lack of Debye screening, implies that electric fields transverse to the applied magnetic field are not modified at all in the color superconducting medium. Possible effects of this anisotropic susceptibility for compact astrophysical objects were discussed in Ref. \cite{EM}. Also, the interplay between the electric properties of dense media and strong magnetic fields could be of interest for future low-temperature/high-density heavy-ion collision experiments where high magnetic and electric fields can be generated \cite{HIC} and presumably CS can be realized \cite{MagneticmomentMCFL, TCS}.

It is important to point out that possibly the situation will be slightly different in the 2SC superconductor in a magnetic field. Even though we expect a similar anisotropy to be present in the 2SC case too, the charged blue quarks can in principle Debye screens there an external electric field in all directions. It will be worth investigating this case in more detail since 2SC superconductivity can be the more reliable phase at moderate densities and low temperatures \cite{Strong-Coup.}.

\begin{acknowledgments}
This work has been supported in part by DOE Nuclear Theory grant DE-SC0002179. The authors want to thank C. Manuel, H-c Ren, A. Sanchez and I.A. Shovkovy for helpful discussions.
\end{acknowledgments}

\appendix
\section{Transversality of the rotated photon polarization tensor in the strong-field Approximation}

The rotated electrodynamics in a color superconducting medium is a $\widetilde{U}(1)$ gauge theory. As a consequence, the in-medium photon self-energy operator should be transverse with respect to photon four-momentum. In the strong-field approximation, when all the quarks are confined to the LLL, the photon self-energy tensor is reduced to the (0,1) plane of the longitudinal-momentum (1+1)-D space, as has been shown earlier. Although from the covariant structure of $\Pi_{\mu\nu}$ (\ref{Cov-structure}), the transversality of the in-medium photon self energy is already guaranteed, we want to show in this Appendix that the transversality is intimately associated in this case to the structure of the quark propagator in the LLL. Also, we want to highlight the relevance of the Pauli-Villars regularization in securing the transversality.

In the strong-field approximation,where the quarks are confined to the LLL, the inverse quark propagator is given by
\begin{equation}
[{\tilde {\cal S}}^{l=0}_{(+)}]^{-1}(k^\parallel)={\cal S}^{-1}(k^\parallel)={\slashed k}^\parallel+\mu\gamma_0\sigma_3+i\gamma_5\Delta_0\Delta(+)\sigma_2,
\label{Inv-Grenn-Funct}
\end{equation}
where the Pauli matrices, $\sigma_i$, act on the Nambu-Gorkov space. With the same compact notation, the vertex (\ref{vetex}) for the positively charged particles is given by the diagonal matrix in Nambu-Gorkov space
\begin{equation}
\Gamma^\mu_{AB}={\tilde e}\gamma^\mu \delta_{AB}.
\end{equation}
Where $A,B$ are Nambu-Gorkov indices. From (\ref{Inv-Grenn-Funct}) we have
\begin{equation}
{\slashed q}^\parallel={\cal S}^{-1}(k^\parallel+q^\parallel)-{\cal S}^{-1}(k^\parallel),
\label{vetexinNG}
\end{equation}

Sandwiching (\ref{vetexinNG}) between ${\cal S}(k^\parallel+q^\parallel)$ and ${\cal S}(k^\parallel)$, we find that
\begin{equation}
{\cal S}(k^\parallel+q^\parallel){\slashed q}^\parallel{\cal S}(k^\parallel)={\cal S}(k^\parallel)-{\cal S}(k^\parallel+q^\parallel).
\label{sanwichingvertex}
\end{equation}

The transversality of the rotated photon self-energy tensor requires $(q^\mu)^\parallel{\Pi_{\mu\nu}^\parallel}$, which from (\ref{polarizationtensor}) is equivalent to
\begin{equation}
T\sum_{q_0}\int\frac{dq_3}{(2\pi)^2}{\rm Tr}\left[{\cal S}(k^\parallel+q^\parallel)\Delta(+){\tilde e}{\slashed q}^\parallel{\cal S}(k^\parallel)\Delta(+){\tilde e}\gamma_\nu^\parallel\right]=0,
\label{gaugedtensor}
\end{equation}
From the fact that $[\Delta(+), {\cal S}(k^\parallel)]=0$, it follows from (\ref{sanwichingvertex}) and (\ref{gaugedtensor}) that
\begin{align}
T\sum_{q_0}\int\frac{dq_3}{(2\pi)^2} \left\{{\rm Tr}\left[\Delta(+){\cal S}(k^\parallel){\tilde e}\gamma_\nu^\parallel\right]\right.-\left.{\rm Tr}\left[\Delta(+){\cal S}(k^\parallel+q^\parallel){\tilde e}\gamma_\nu^\parallel\right]\right\}=0.
\end{align}
 Naively, this requirement is satisfied by simply shifting the momentum. However, the shift of the integration variable is no longer legitimate if there is a linear or higher divergence. We thus introduce the Pauli-Villars regularization and the regularized version of the LHS of (\ref{gaugedtensor}) reads
\begin{align}
T\sum_{q_0}\int\frac{dq_3}{(2\pi)^2}\left\{{\rm Tr}\left[{\cal S}(k^\parallel+q^\parallel)\Delta(+){\tilde e}{\slashed q}^\parallel{\cal S}(k^\parallel)\Delta(+){\tilde e}\gamma_\nu^\parallel\right]\right.-\left.{\rm Tr}\left[{\cal S}_\Lambda (k^\parallel+q^\parallel)\Delta(+){\tilde e}{\slashed q}^\parallel{\cal S}_\Lambda (k^\parallel)\Delta(+){\tilde e}\gamma_\nu^\parallel\right]\right\}.
\end{align}
where ${\cal S}_\Lambda (k)$ is the Pauli-Villars propagator with zero gap, zero chemical potential and mass $\Lambda$. We observe that ${\cal S}_\Lambda(k)$ satisfies the same relation (\ref{sanwichingvertex}). Therefore
\begin{equation}
(q^\mu)^\parallel{\Pi_{\mu\nu}^\parallel}_R=\frac{{\tilde e}^2}{2}({\tilde e}{\tilde B})T\sum_{q_0}\int\frac{dq_3}{(2\pi)^2}\left[F_\nu(k^\parallel)-F_\nu(k^\parallel+q^\parallel)\right],
\label{regulariedtensor}
\end{equation}
where
\begin{equation}
F_\nu(k)={\rm Tr}\left[\Delta(+){\cal S}(k^\parallel){\tilde e}\gamma_\nu^\parallel\right]-{\rm Tr}\left[\Delta(+){\cal S}_\Lambda(k^\parallel){\tilde e}\gamma_\nu^\parallel\right].
\end{equation}
It can be shown explicitly that for a fixed cutoff $\Lambda$, neither of the integrals in (\ref{regulariedtensor}) is more divergent than logarithmically. Then, the shift of the integration variable is legitimate and we thus obtain
\begin{equation}
(q^\mu)^\parallel{\Pi_{\mu\nu}^\parallel}_R=0.
\end{equation}
as required by gauge invariance.


\end{document}